# Quantum entanglement dynamics in a three-qubit system interacting with a spin chain

Seyed Mohsen Moosavi Khansari[1]
*Department of Physics, Faculty of Basic Sciences, Ayatollah Boroujerdi University, Boroujerd, IRAN.*
Fazlollah Kazemi Hasanvand[2]
*Department of Physics, Faculty of Basic Sciences, Ayatollah Boroujerdi University, Boroujerd, IRAN.*
Nader Habibi[3]
*Department of Mathematics, Faculty of Basic Sciences, Ayatollah Boroujerdi University, Boroujerd, IRAN.*

This paper investigates the evolution of the entanglement of three-qubit states in a spin-chain nvironment. Utilizing negativity as a metric for entanglement assessment, we focus on the $GHZ$ (Greenberger-Horne-Zeilinger state), $W$ (Wolfgang Dur state), and $W_\xi$ (A Wolfgang Dur state) quantum states as the initial system states. We conducted a thorough exploration and detailed analysis of the entanglement dynamics exhibited by these specific states, focusing on various parameters that nfluence their behavior over time. Our results closely align with findings from similar studies, reinforcing the validity of our conclusions and enhancing our understanding of quantum entanglement dynamics in complex systems.

## 1 Introduction

Entanglement is a key aspect of quantum mechanics that has attracted significant attention recently. The applicability of the proposed method in calculations and quantum information is noteworthy. The impossibility of completely isolating a quantum system underscores its significance. Investigating the entanglement behavior of the system's initial state in the presence of environmental effects is crucial. The interaction between a quantum system and its environment gives rise to decoherence, a phenomenon that can weaken or even terminate entanglement [1, 2, 3, 8]. This paper delves into the Hamiltonian of the system-environment setup, which comprises noninteracting qubits and a coupled spin chain. We explore the initial $GHZ$ and $W$ and $W_\xi$ quantum states by evaluating the entanglement dynamics of the system environment using the negativity measure. In addition, we examined how these dynamics vary for different parameters.

The technical motivation driving the exploration of quantum entanglement dynamics within a three-qubit system is fundamentally rooted in the vast range of potential applications that this line of research offers in quantum computing and quantum information theory. One key aspect is the complexity and diversity associated with three-qubit systems. A three-qubit system presents a much richer tapestry of entanglement patterns and interactions than single- or two-qubit systems, which are relatively simpler and have limited entanglement configurations. This increased complexity is essential for deepening our understanding of fundamental quantum phenomena, allowing researchers to uncover new insights and deepen our understanding of quantum mechanics. In addition, the investigation of entanglement dynamics plays a significant role in the development of quantum algorithms. By gaining a thorough understanding of how entanglement evolves, researchers can design and implement more efficient quantum algorithms. This advancement can substantially enhance computational power, enabling solutions to specific problems that would be impossible to tackle using classical computation methods, thus showcasing the true power of quantum computing. The investigation into the entanglement dynamics is vital for designing error correction mechanisms. Specifically, this research helps formulate fault-tolerant quantum error correction codes. Such codes are indispensable for ensuring the reliability and stability of quantum computing systems, which are often susceptible to various errors due to decoherence and other external factors. Quantum communication is another area that can benefit from a deeper understanding of three-qubit entanglement. The insights gained from studying entangled states involving three qubits can lead to quantum communication protocol improvements. For instance, this knowledge can enhance secure information transmission methods, such as quantum key distribution, which is crucial for developing secure communication channels based on the principles of quantum mechanics. Lastly, exploring these intricate systems provides a valuable avenue for testing and verifying fundamental quantum mechanics theories. Studying three-qubit entanglement allows researchers to explore the limits of entanglement and examine concepts like non-locality. These investigations hold profound implications for our broader understanding of the foundational principles underlying quantum mechanics and how they relate to the nature of reality itself. In summary, the study of quantum entanglement dynamics in three-qubit systems represents a vital and impactful area of research. It not only holds significant theoretical importance but also promises a wide array of practical applications that could shape the future of technology and our comprehension of the quantum world.

---

[1] Email (corresponding author): m.moosavikhansari@abru.ac.ir

[2] Email: fa_kazemi270@yahoo.com

[3] Email: habibi@abru.ac.ir

This study explores the entanglement evolution of three-qubit states interacting with a spin chain environment using negativity as a metric for evaluating entanglement. Because negativity is a parameter used to analyze entanglement dynamics, all conceivable explanations of negativity are valid for this purpose. We now examine the dynamics of quantum entanglement for $GHZ$, $W$, and $W_\xi$ states, based on the proposed Hamiltonian. The results show a significant alignment with findings from similar studies.

This study focuses on analyzing the entanglement dynamics of a qubit system in an environment. The qubits do not interact with each other but interact with the environment, which is itself influenced by a constant magnetic field. The remainder of this paper is structured as follows: Section 2 introduces the model and derives the time-dependent density. Sections 3, 4, 5, and 6 present our initial $GHZ$ state, $W$ state, and $W_\xi$ state, along with their corresponding time-dependent density matrices and the measures outlined in Section 2. Finally, Section 7 presents the conclusions and discussions.

## 2 Theoretical calculations

The system comprises three noninteracting qubits that interact with a spin chain. The total Hamiltonian is denoted as

$$H = S_A^z \sum_{k=1}^{\mathcal{N}} g^A S_{E,k}^z + S_B^z \sum_{k=1}^{\mathcal{N}} g^B S_{E,k}^z + S_C^z \sum_{k=1}^{\mathcal{N}} g^C S_{E,k}^z + \sum_{k=1}^{\mathcal{N}} h_k S_{E,k}^x \quad (1)$$

In quantum mechanics, the Hamiltonian is a mathematical operator that describes the total energy of a given system. In this particular case, $S_A^z$, $S_B^z$ and $S_C^z$ are the spin 1/2 operators in the z direction for the three qubits, respectively. Meanwhile, $S_{E,k}^z$ and $S_{E,k}^x$ are the spin 1/2 operators in the z direction and the x direction, respectively, for the particles in the spin chain of the environment. The coupling constants of qubits A, B, and C with this spin chain are defined as $g^A$, $g^B$ and $g^C$, respectively. These constants determine the strength of the interaction between the qubits and the particles in the spin chain. $h_k$ is the intensity of the external magnetic field applied to the particles in the spin chain of the environment. But, $\mathcal{N}$ denotes the number of particles in the spin chain. The Hamiltonian accounts for the interactions between the qubits and the environment and is given by the sum of the coupling constants multiplied by the tensor products of the spin operators. Overall, these mathematical operators and constants provide a detailed description of the energy of the quantum mechanical system under investigation.

The initial state of the system is a three-qubit pure state represented by

$$|\psi_s(0)\rangle = C_1|000\rangle + C_2|001\rangle + C_3|010\rangle + C_4|011\rangle + C_5|100\rangle + C_6|101\rangle + C_7|110\rangle + C_8|111\rangle \quad (2)$$

where the coefficients $C_i$ with $i = 1,2,\ldots,8$ must satisfy the following normalization condition:

$$\sum_{i=1}^{8} |C_i|^2 = 1 \quad (3)$$

and $|000\rangle$ represents the three-qubit system as $|0_A\rangle \otimes |0_B\rangle \otimes |0_C\rangle$, with the remaining states expressed similarly. We define the initial state of the environment as

$$|\psi_E(0)\rangle = \prod_{i=1}^{\mathcal{N}} (\gamma_i|0\rangle + \eta_i|1\rangle) \quad (4)$$

where the normalization condition $|\gamma_i|^2 + |\eta_i|^2 = 1$ is required, with $\gamma_i = \eta_i = 1/\sqrt{2}$. We express the system's density operator, obtained by tracing out the environment, as follows:

$$\rho_s(t) = tr_E\big[e^{-iHt}|\psi_s(0)\rangle\langle\psi_s(0)| \otimes |\psi_E(0)\rangle\langle\psi_E(0)|e^{iHt}\big] \quad (5)$$

In the three-qubit computational bases $|000\rangle,\ldots,|111\rangle$, the matrix elements $\rho_s(t)$ are defined as follows:

$$M_{\alpha\beta} = C_\alpha C_\beta^* F_{\alpha\beta}, \quad \alpha,\beta = 1,2,\ldots,8 \quad (6)$$

where $F_{\alpha\beta}$, known as decoherence factors, are defined as follows:

$$\big[F_{\alpha,\beta}^{\mathcal{N}}\big]_{\mathcal{N}=1} = (\gamma_i)^*\eta_i \times$$

$$+\gamma_i(\eta_i)^* \left( \frac{ih_i\sin(t\Lambda_\beta)\left(\cos(t\Lambda_\alpha) + \frac{iA_\alpha\sin(t\Lambda_\alpha)}{\Lambda_\alpha}\right)}{\Lambda_\beta} - \frac{ih_i\sin(t\Lambda_\alpha)\left(\cos(t\Lambda_\beta) + \frac{iA_\beta\sin(t\Lambda_\beta)}{\Lambda_\beta}\right)}{\Lambda_\alpha} \right)$$

$$+\gamma_i(\eta_i)^* \left( \frac{ih_i\sin(t\Lambda_\beta)\left(\cos(t\Lambda_\alpha) - \frac{iA_\alpha\sin(t\Lambda_\alpha)}{\Lambda_\alpha}\right)}{\Lambda_\beta} + \frac{ih_i\sin(t\Lambda_\alpha)\left(-\cos(t\Lambda_\beta) + \frac{iA_\beta\sin(t\Lambda_\beta)}{\Lambda_\beta}\right)}{\Lambda_\alpha} \right)$$

$$+\gamma_i(\gamma_i)^* \left(\cos(t\Lambda_\alpha) - \frac{iA_\alpha\sin(t\Lambda_\alpha)}{\Lambda_\alpha}\right)\left(\cos(t\Lambda_\beta) + \frac{iA_\beta\sin(t\Lambda_\beta)}{\Lambda_\beta}\right) + \eta_i(\eta_i)^* \times$$

$$\left(\cos(t\Lambda_\alpha) + \frac{iA_\alpha\sin(t\Lambda_\alpha)}{\Lambda_\alpha}\right)\left(\cos(t\Lambda_\beta) - \frac{iA_\beta\sin(t\Lambda_\beta)}{\Lambda_\beta}\right) + \frac{h_i^2\sin(t\Lambda_\alpha)\sin(t\Lambda_\beta)}{\Lambda_\alpha\Lambda_\beta} \quad (7)$$

In this mathematical relationship, the following definitions are used:

$$\Lambda_{\alpha(\beta)} = \sqrt{\zeta_{\alpha(\beta)}^2 + h_i^2} \quad (8)$$

$$\zeta_1 = \frac{1}{2}(g^A + g^B + g^C) \quad (9)$$

$$\zeta_2 = \frac{1}{2}(g^A + g^B - g^C) \quad (10)$$

$$\zeta_3 = \frac{1}{2}(g^A - g^B + g^C) \quad (11)$$

$$\zeta_4 = \frac{1}{2}(g^A - g^B - g^C) \quad (12)$$

$$\zeta_5 = \frac{1}{2}(-g^A + g^B + g^C) \quad (13)$$

$$\zeta_6 = \frac{1}{2}(-g^A + g^B - g^C) \quad (14)$$

$$\zeta_7 = \frac{1}{2}(-g^A - g^B + g^C) \quad (15)$$

$$\zeta_8 = \frac{1}{2}(-g^A - g^B - g^C) \quad (16)$$

Equation (8) is a straightforward substitution. Equations (9) to (16) describe the Hamiltonian effect on the three-qubit state, specifically, the eigenvalues corresponding to its eigenvectors.

Quantum negativity is a measure of quantum entanglement in the mixed state of a bipartite quantum system. This quantifies the degree to which a quantum state's partial transpose is not a positive semidefinite operator. When the partial transpose of a density matrix has negative eigenvalues, the negativity value reflects the extent of the entanglement. This concept is particularly useful for analyzing quantum correlations beyond the classical limits and can help distinguish between different quantum states. Quantum negativity is a measure of quantum entanglement that quantifies the degree to which a quantum state deviates from being separable. It is particularly useful for assessing entanglement in mixed states. There are some common methods for calculating quantum negativity: The first step involves taking the partial transpose of the density matrix of the quantum state concerning one of the subsystems. For a bipartite system described by a density matrix $\rho$, the partial transpose concerning subsystem $A$ is denoted as $\rho^{T_A}$. Eigenvalue Calculation: After obtaining the partially transposed density matrix, the next step is to compute its eigenvalues. The eigenvalues can be found using numerical methods or analytical techniques, depending on the complexity of the density matrix.

The negativity of a quantum state with the density matrix $\rho$ is defined as follows [14]:

$$N(\rho) = \frac{||\rho^{T_i}||-1}{2} \quad (17)$$

In this framework, $\rho^{T_i}$ denotes the partial transpose of $\rho$ regarding the specific component identified as $i$. If $N > 0$, the state is entangled and if $N = 0$, the state is separable. The larger the negativity, the stronger is the entanglement. For complex systems, numerical techniques such as Monte Carlo Methods or optimization algorithms can be employed to estimate the negativity, especially when dealing with large-density matrices. These methods provide a systematic approach to quantifying quantum entanglement in various quantum systems, thus improving our understanding of quantum correlations. We use the negativity measure to compute entanglement.

The numerical methods used to calculate negativity in the discussion of quantum entanglement are summarized as follows:
- For complex systems, Monte Carlo techniques may be employed to sample different states and estimate the average negativity. This method can be useful for larger systems where exact diagonalization becomes computationally expensive.
- In systems where the Hamiltonian is complex, variational methods can help find approximate ground states, which can then be used to compute the density matrix and its negativity.
- For lattice systems, tensor network approaches like Matrix Product States (MPS) can efficiently represent and compute entanglement properties, including negativity.
- Computational Tools: Numerical calculation software packages such as Python (using libraries like NumPy and QuTiP), MATLAB, Mathematica and specialized quantum computing frameworks can help implement these methods.

## 3 The GHZ state is the initial state of the three-qubit system

The $GHZ$ (Greenberger-Horne-Zeilinger) state is an entangled quantum state involving multiple particles that is typically used in quantum information theory and quantum mechanics to illustrate the phenomenon of quantum entanglement. For $n$ qubits, the $GHZ$ state is defined as follows:

$$|\text{GHZ}\rangle = \frac{1}{\sqrt{2}}\left(|0\rangle^{\otimes n} + |1\rangle^{\otimes n}\right) \quad (18)$$

For $n = 3$, we consider the initial state of the system as $GHZ$ state with the following relationship:

$$|\text{GHZ}\rangle = \frac{1}{\sqrt{2}}(|000\rangle + |111\rangle) \quad (19)$$

The initial ($t = 0$) density matrix of the system corresponding to this state is written as follows:

$$[\rho_s(0)]_{GHZ} = |\text{GHZ}\rangle\langle\text{GHZ}| =$$

$$\begin{pmatrix}
\frac{1}{2} & 0 & 0 & 0 & 0 & 0 & 0 & \frac{1}{2} \\
0 & 0 & 0 & 0 & 0 & 0 & 0 & 0 \\
0 & 0 & 0 & 0 & 0 & 0 & 0 & 0 \\
0 & 0 & 0 & 0 & 0 & 0 & 0 & 0 \\
0 & 0 & 0 & 0 & 0 & 0 & 0 & 0 \\
0 & 0 & 0 & 0 & 0 & 0 & 0 & 0 \\
0 & 0 & 0 & 0 & 0 & 0 & 0 & 0 \\
\frac{1}{2} & 0 & 0 & 0 & 0 & 0 & 0 & \frac{1}{2}
\end{pmatrix} \quad (20)$$

The Dirac symbol, also known as Dirac notation or bra-ket notation, is a standard notation in quantum mechanics that is used to describe quantum states and their properties. On the other hand, after interacting with the environment, the density matrix of the system is expressed as follows:

$$(\rho_s(t))_{GHZ} = \frac{1}{2}(F_{18}^{*}|111\rangle\langle 000| + F_{18}|000\rangle\langle 111| + |000\rangle\langle 000| + |111\rangle\langle 111|) \quad (21)$$

$F_{18}$ is a decoherence factor, and $F_{18}^{*}$ is its complex conjugate. After some algebraic calculations, it can be rewritten as a matrix with eight bases of the three corresponding qubits as follows:

$$[\rho_s(t)]_{GHZ} =$$

$$\begin{pmatrix}
\frac{1}{2} & 0 & 0 & 0 & 0 & 0 & 0 & \frac{F_{18}}{2} \\
0 & 0 & 0 & 0 & 0 & 0 & 0 & 0 \\
0 & 0 & 0 & 0 & 0 & 0 & 0 & 0 \\
0 & 0 & 0 & 0 & 0 & 0 & 0 & 0 \\
0 & 0 & 0 & 0 & 0 & 0 & 0 & 0 \\
0 & 0 & 0 & 0 & 0 & 0 & 0 & 0 \\
0 & 0 & 0 & 0 & 0 & 0 & 0 & 0 \\
\frac{F_{18}^{*}}{2} & 0 & 0 & 0 & 0 & 0 & 0 & \frac{1}{2}
\end{pmatrix} \quad (22)$$

First, the calculations were focused solely on the $A$ subsystem. The partial transpose of this subsystem can be expressed as follows:

$$[\rho_s(t)]_{GHZ}^{T_A} =$$

$$\begin{pmatrix} \frac{1}{2} & 0 & 0 & 0 & 0 & 0 & 0 & 0 \\ 0 & 0 & 0 & 0 & 0 & 0 & 0 & 0 \\ 0 & 0 & 0 & 0 & 0 & 0 & 0 & 0 \\ 0 & 0 & 0 & 0 & \frac{1}{2}F_{18}^* & 0 & 0 & 0 \\ 0 & 0 & 0 & \frac{1}{2}F_{18} & 0 & 0 & 0 & 0 \\ 0 & 0 & 0 & 0 & 0 & 0 & 0 & 0 \\ 0 & 0 & 0 & 0 & 0 & 0 & 0 & 0 \\ 0 & 0 & 0 & 0 & 0 & 0 & 0 & \frac{1}{2} \end{pmatrix} \quad (23)$$

This matrix is Hermitian; thus, its trace norm is equal to the sum of the absolute values of its eigenvalues. By the negativity relation, we can express this as

$$N_{A-BC} = \frac{1+|F_{18}|^N - 1}{2} = \frac{|F_{18}|^N}{2} \quad (24)$$

Using similar calculations, we derive the following relationship for the partial transpose of the subsystem $B$:

$$[\rho_s(t)]_{GHZ}^{T_B} =$$

$$\begin{pmatrix} \frac{1}{2} & 0 & 0 & 0 & 0 & 0 & 0 & 0 \\ 0 & 0 & 0 & 0 & 0 & 0 & 0 & 0 \\ 0 & 0 & 0 & 0 & 0 & \frac{F_{18}}{2} & 0 & 0 \\ 0 & 0 & 0 & 0 & 0 & 0 & 0 & 0 \\ 0 & 0 & 0 & 0 & 0 & 0 & 0 & 0 \\ 0 & 0 & \frac{F_{18}^*}{2} & 0 & 0 & 0 & 0 & 0 \\ 0 & 0 & 0 & 0 & 0 & 0 & 0 & 0 \\ 0 & 0 & 0 & 0 & 0 & 0 & 0 & \frac{1}{2} \end{pmatrix} \quad (25)$$

This matrix is also Hermitian, and its trace norm is equal to the sum of the absolute values of its eigenvalues. Therefore, based on the negativity relation, we can express it as follows:

$$N_{B-CA} = \frac{1+|F_{18}|^N - 1}{2} = \frac{|F_{18}|^N}{2} \quad (26)$$

We can represent the relationship for the partial transpose of subsystem $C$ as follows:

$$[\rho_s(t)]_{GHZ}^{T_C} =$$

$$\begin{pmatrix} \frac{1}{2} & 0 & 0 & 0 & 0 & 0 & 0 & 0 \\ 0 & 0 & 0 & 0 & 0 & 0 & \frac{F_{18}}{2} & 0 \\ 0 & 0 & 0 & 0 & 0 & 0 & 0 & 0 \\ 0 & 0 & 0 & 0 & 0 & 0 & 0 & 0 \\ 0 & 0 & 0 & 0 & 0 & 0 & 0 & 0 \\ 0 & 0 & 0 & 0 & 0 & 0 & 0 & 0 \\ 0 & \frac{F_{18}^*}{2} & 0 & 0 & 0 & 0 & 0 & 0 \\ 0 & 0 & 0 & 0 & 0 & 0 & 0 & \frac{1}{2} \end{pmatrix} \quad (27)$$

It can be demonstrated that this matrix is also Hermitian, and its trace norm is equal to the sum of the absolute values of its eigenvalues. Therefore, according to the negativity relation, we can express the following:

$$N_{C-AB} = \frac{1+|F_{18}|^N - 1}{2} = \frac{|F_{18}|^N}{2} \quad (28)$$

## 4 The W state is the initial state of the three-qubit system

The quantum $W$ state is a multipartite entangled state that generalizes the concept of entanglement beyond the well-known Bell states or $GHZ$ states. The $W$ state is particularly significant for its resistance to loss of entanglement when some particles are measured or discarded. For $n$ qubits, the $W$ state can be expressed as

$$|W\rangle = \frac{1}{\sqrt{n}}(|000\ldots 1\rangle + \ldots + |010\ldots 0\rangle + |100\ldots 0\rangle) \quad (29)$$

For $n = 3$, we define the state $W$ as the initial state of the system as follows:

$$|W\rangle = \frac{1}{\sqrt{3}}(|001\rangle + |010\rangle + |100\rangle) \quad (30)$$

The initial density matrix of the system corresponding to this state is as follows:

$$[\rho_s(0)]_W = |W\rangle\langle W| =$$

$$\begin{pmatrix} 0 & 0 & 0 & 0 & 0 & 0 & 0 & 0 \\ 0 & \frac{1}{3} & \frac{1}{3} & 0 & \frac{1}{3} & 0 & 0 & 0 \\ 0 & \frac{1}{3} & \frac{1}{3} & 0 & \frac{1}{3} & 0 & 0 & 0 \\ 0 & 0 & 0 & 0 & 0 & 0 & 0 & 0 \\ 0 & \frac{1}{3} & \frac{1}{3} & 0 & \frac{1}{3} & 0 & 0 & 0 \\ 0 & 0 & 0 & 0 & 0 & 0 & 0 & 0 \\ 0 & 0 & 0 & 0 & 0 & 0 & 0 & 0 \\ 0 & 0 & 0 & 0 & 0 & 0 & 0 & 0 \end{pmatrix} \quad (31)$$

In addition, the density matrix of the system in the Dirac symbol after interaction with the environment is calculated as follows:

$$(\rho_s(t))_W = \frac{1}{3}(F_{23}{}^*|010\rangle\langle 001| + F_{25}{}^*|100\rangle\langle 001| + F_{35}{}^*|100\rangle\langle 010| + F_{23}|001\rangle\langle 010| + F_{25}|001\rangle\langle 100| + F_{35}|010\rangle\langle 100| + |001\rangle\langle 001| + |010\rangle\langle 010| + |100\rangle\langle 100|) \quad (32)$$

$F_{23}$ is a decoherence factor, and $F_{23}^*$ is its complex conjugate, with similar cases defined analogously. According to the previous computation, after some algebraic calculations, it can be rewritten in the form of a matrix and in eight bases of the corresponding three qubits as follows:

$$[\rho_s(t)]_W =$$

$$\begin{pmatrix} 0 & 0 & 0 & 0 & 0 & 0 & 0 & 0 \\ 0 & \frac{1}{3} & \frac{F_{23}}{3} & 0 & \frac{F_{25}}{3} & 0 & 0 & 0 \\ 0 & \frac{F_{23}{}^*}{3} & \frac{1}{3} & 0 & \frac{F_{35}}{3} & 0 & 0 & 0 \\ 0 & 0 & 0 & 0 & 0 & 0 & 0 & 0 \\ 0 & \frac{F_{25}{}^*}{3} & \frac{F_{35}{}^*}{3} & 0 & \frac{1}{3} & 0 & 0 & 0 \\ 0 & 0 & 0 & 0 & 0 & 0 & 0 & 0 \\ 0 & 0 & 0 & 0 & 0 & 0 & 0 & 0 \\ 0 & 0 & 0 & 0 & 0 & 0 & 0 & 0 \end{pmatrix} \quad (33)$$

Based on the previous scenario, our focus is primarily on subsystem A when performing computations. The partial transpose of this subsystem can be represented as follows:

$$[\rho_s(t)]_W^{T_A} =$$

$$\begin{pmatrix} 0 & 0 & 0 & 0 & 0 & \frac{F_{25}{}^*}{3} & \frac{F_{35}{}^*}{3} & 0 \\ 0 & \frac{1}{3} & \frac{F_{23}}{3} & 0 & 0 & 0 & 0 & 0 \\ 0 & \frac{F_{23}{}^*}{3} & \frac{1}{3} & 0 & 0 & 0 & 0 & 0 \\ 0 & 0 & 0 & 0 & 0 & 0 & 0 & 0 \\ 0 & 0 & 0 & 0 & \frac{1}{3} & 0 & 0 & 0 \\ \frac{F_{25}}{3} & 0 & 0 & 0 & 0 & 0 & 0 & 0 \\ \frac{F_{35}}{3} & 0 & 0 & 0 & 0 & 0 & 0 & 0 \\ 0 & 0 & 0 & 0 & 0 & 0 & 0 & 0 \end{pmatrix} \quad (34)$$

because the matrix is Hermitian, its trace norm is equal to the sum of the absolute values of its eigenvalues. The negativity relation can be expressed as follows:

$$N_{A-BC} = \frac{1}{3}\left(1 + \sqrt{|F_{25}^N|^2 + |F_{35}^N|^2} - 1\right) = \frac{1}{3}\sqrt{|F_{25}^N|^2 + |F_{35}^N|^2} \quad (35)$$

For this state, we derive the following relationship for the partial transpose of subsystem B:

$$[\rho_s(t)]_W^{T_B} = \begin{pmatrix} 0 & 0 & 0 & \frac{F_{23}^*}{3} & 0 & 0 & \frac{F_{35}}{3} & 0 \\ 0 & \frac{1}{3} & 0 & 0 & \frac{F_{25}}{3} & 0 & 0 & 0 \\ 0 & 0 & \frac{1}{3} & 0 & 0 & 0 & 0 & 0 \\ \frac{F_{23}}{3} & 0 & 0 & 0 & 0 & 0 & 0 & 0 \\ 0 & \frac{F_{25}^*}{3} & 0 & 0 & \frac{1}{3} & 0 & 0 & 0 \\ 0 & 0 & 0 & 0 & 0 & 0 & 0 & 0 \\ \frac{F_{35}^*}{3} & 0 & 0 & 0 & 0 & 0 & 0 & 0 \\ 0 & 0 & 0 & 0 & 0 & 0 & 0 & 0 \end{pmatrix} \quad (36)$$

This matrix is also Hermitian. The trace norm is equal to the sum of the absolute values of its eigenvalues. According to the negativity relation, we can derive the following:

$$N_{B-CA} = \frac{1}{2}\left(1 + \frac{2}{3}\sqrt{|F_{23}^N|^2 + |F_{35}^N|^2} - 1\right) = \frac{1}{3}\sqrt{|F_{23}^N|^2 + |F_{35}^N|^2} \quad (37)$$

For the partial transposition of subsystem C, the following relation can be derived:

$$[\rho_s(t)]_W^{T_C} = \begin{pmatrix} 0 & 0 & 0 & \frac{F_{23}}{3} & 0 & \frac{F_{25}}{3} & 0 & 0 \\ 0 & \frac{1}{3} & 0 & 0 & 0 & 0 & 0 & 0 \\ 0 & 0 & \frac{1}{3} & 0 & \frac{F_{35}}{3} & 0 & 0 & 0 \\ \frac{F_{23}^*}{3} & 0 & 0 & 0 & 0 & 0 & 0 & 0 \\ 0 & 0 & \frac{F_{35}^*}{3} & 0 & \frac{1}{3} & 0 & 0 & 0 \\ \frac{F_{25}^*}{3} & 0 & 0 & 0 & 0 & 0 & 0 & 0 \\ 0 & 0 & 0 & 0 & 0 & 0 & 0 & 0 \\ 0 & 0 & 0 & 0 & 0 & 0 & 0 & 0 \end{pmatrix} \quad (38)$$

This matrix is also Hermitian, and its trace norm is equal to the sum of the absolute values of its eigenvalues. By the negativity relation, we can express:

$$N_{C-AB} = \frac{1}{2}\left(1 + \frac{2}{3}\sqrt{|F_{23}^{\mathcal{N}}|^2 + |F_{25}^{\mathcal{N}}|^2} - 1\right) = \frac{1}{3}\sqrt{|F_{23}^{\mathcal{N}}|^2 + |F_{25}^{\mathcal{N}}|^2} \quad (39)$$

## 5 The $W_\xi$ state serves as the initial state of the three-qubit system

Finally, we define state $W_\xi$ as the system's initial state as follows:

$$|W_\xi\rangle = \frac{1}{\sqrt{2\xi+2}}\left(e^{i\delta}\sqrt{\xi+1}|001\rangle + e^{i\phi}\sqrt{\xi}|010\rangle + |100\rangle\right) \quad (40)$$

where $\xi$ is a natural number and $\delta$ and $\phi$ are real numbers. For this state the non-zero elements of the initial density matrix, $[\rho_s(0)]_{W_\xi} = |W_\xi\rangle\langle W_\xi|$, are defined as follows:

$$[\rho_s(0)]_{W_\xi}^{(2,2)} = 1/2, \quad [\rho_s(0)]_{W_\xi}^{(2,3)} = \frac{\xi e^{i(\delta-\phi)}}{2\sqrt{\xi(\xi+1)}}$$

$$[\rho_s(0)]_{W_\xi}^{(2,5)} = \frac{e^{i\delta}}{2\sqrt{\xi+1}}, \quad [\rho_s(0)]_{W_\xi}^{(3,2)} = \frac{\xi e^{-i(\delta-\phi)}}{2\sqrt{\xi(\xi+1)}}$$

$$[\rho_s(0)]_{W_\xi}^{(3,3)} = \frac{\xi}{2\xi+2}, \quad [\rho_s(0)]_{W_\xi}^{(3,5)} = \frac{\sqrt{\xi}e^{i\phi}}{2\xi+2}$$

$$[\rho_s(0)]_{W_\xi}^{(5,2)} = \frac{e^{-i\delta}}{2\sqrt{\xi+1}}, \quad [\rho_s(0)]_{W_\xi}^{(5,3)} = \frac{\sqrt{\xi}e^{-i\phi}}{2\xi+2}$$

$$[\rho_s(0)]_{W_\xi}^{(5,5)} = \frac{1}{2\xi+2} \quad (41)$$

After performing some algebraic calculations, the non-zero elements of the density matrix of the system, $[\rho_s(t)]_{W_\xi}$, following interaction with the environment for this state can be expressed as a matrix and about the eight bases of the corresponding three qubits as follows:

$$[\rho_s(t)]_{W_\xi}^{(2,2)} = 1/2, \quad [\rho_s(t)]_{W_\xi}^{(2,3)} = \frac{F_{23}\xi e^{i(\delta-\phi)}}{2\sqrt{\xi(\xi+1)}}$$

$$[\rho_s(t)]_{W_\xi}^{(2,5)} = \frac{e^{i\delta}F_{25}}{2\sqrt{\xi+1}}, \quad [\rho_s(t)]_{W_\xi}^{(3,2)} = \frac{\xi F_{23}^* e^{-i(\delta-\phi)}}{2\sqrt{\xi(\xi+1)}}$$

$$[\rho_s(t)]_{W_\xi}^{(3,3)} = \frac{\xi}{2\xi+2}, \quad [\rho_s(t)]_{W_\xi}^{(3,5)} = \frac{F_{35}\sqrt{\xi}e^{i\phi}}{2\xi+2}$$

$$[\rho_s(t)]_{W_\xi}^{(5,2)} = \frac{e^{-i\delta}F_{25}^*}{2\sqrt{\xi+1}}, \quad [\rho_s(t)]_{W_\xi}^{(5,3)} = \frac{\sqrt{\xi}e^{-i\phi}F_{35}^*}{2\xi+2}$$

$$[\rho_s(t)]_{W_\xi}^{(5,5)} = \frac{1}{2\xi+2} \quad (42)$$

In this state, the calculations were initially concentrated solely on subsystem A. For this subsystem, the non-zero elements of the density matrix's partial transpose can be expressed as follows:

$$[\rho_s(t)]_{W_\xi}^{T_A\,(1,6)} = \frac{e^{-i\delta}F_{25}^*}{2\sqrt{\xi+1}}, [\rho_s(t)]_{W_\xi}^{T_A\,(1,7)} = \frac{\sqrt{\xi}e^{-i\phi}F_{35}^*}{2\xi+2}$$

$$[\rho_s(t)]_{W_\xi}^{T_A\,(2,2)} = \frac{1}{2}, [\rho_s(t)]_{W_\xi}^{T_A\,(2,3)} = \frac{F_{23}\xi e^{i(\delta-\phi)}}{2\sqrt{\xi(\xi+1)}}$$

$$[\rho_s(t)]_{W_\xi}^{T_A\,(3,2)} = \frac{\xi F_{23}^* e^{-i(\delta-\phi)}}{2\sqrt{\xi(\xi+1)}}, [\rho_s(t)]_{W_\xi}^{T_A\,(3,3)} = \frac{\xi}{2\xi+2}$$

$$[\rho_s(t)]_{W_\xi}^{T_A\,(5,5)} = \frac{1}{2\xi+2}, [\rho_s(t)]_{W_\xi}^{T_A\,(6,1)} = \frac{e^{i\delta}F_{25}}{2\sqrt{\xi+1}}$$

$$[\rho_s(t)]_{W_\xi}^{T_A\,(7,1)} = \frac{F_{35}\sqrt{\xi}e^{i\phi}}{2\xi+2} \quad (43)$$

This matrix is Hermitian; thus, its trace norm is equal to the sum of the absolute values of the eigenvalues of this matrix. According to this, the negativity relation can be written as follows:

$$N_{A-BC} = -\frac{1}{8(\xi+1)^{5/2}\sqrt{\xi(\xi+1)}}$$

$$\left(-\xi\left|\sqrt{\xi(\xi+1)}(2\xi+1) - e^{-i(\delta+\phi)}\sqrt{e^{2i(\delta+\phi)}\xi(\xi+1)^2\left(4\xi(\xi+1)|F_{23}^\mathcal{N}|^2 + 1\right)}\right|\right.$$

$$-\left|\sqrt{\hat{\iota}(\xi+1)}(2\xi+1) - e^{-i(\delta+\phi)}\sqrt{e^{2i(\delta+\phi)}\xi(\xi+1)^2\left(4\xi(\xi+1)|F_{23}^\mathcal{N}|^2 + 1\right)}\right|$$

$$-\xi\left|\sqrt{\xi(\xi+1)}(2\xi+1) + e^{-i(\delta+\phi)}\sqrt{e^{2i(\delta+\phi)}\xi(\xi+1)^2\left(4\xi(\xi+1)|F_{23}^\mathcal{N}|^2 + 1\right)}\right|$$

$$-\left|\sqrt{\xi(\xi+1)}(2\xi+1) + e^{-i(\delta+\phi)}\sqrt{e^{2i(\delta+\phi)}\xi(\xi+1)^2\left(4\xi(\xi+1)|F_{23}^\mathcal{N}|^2 + 1\right)}\right|$$

$$-4\sqrt{\xi(\xi+1)}(\xi+1)^{3/2}\left|\sqrt{(\xi+1)|F_{25}^\mathcal{N}|^2 + \xi|F_{35}^\mathcal{N}|^2}\right| + 4\xi\sqrt{\xi(\xi+1)}(\xi+1)^{3/2}$$

$$\left. +2\sqrt{\xi(\xi+1)}(\xi+1)^{3/2}\right) \quad (44)$$

Similar to subsystem $A$ for subsystem $B$, the non-zero elements of the partial transpose density matrix can be represented as follows:

$$[\rho_s(t)]_{W_\xi}^{T_B\ (1,4)} = \frac{\xi F_{23}{}^* e^{-i(\delta-\phi)}}{2\sqrt{\xi(\xi+1)}}, [\rho_s(t)]_{W_\xi}^{T_B\ (1,7)} = \frac{F_{35}\sqrt{\xi}e^{i\phi}}{2\xi+2}$$

$$[\rho_s(t)]_{W_\xi}^{T_B\ (2,2)} = 1/2, [\rho_s(t)]_{W_\xi}^{T_B\ (2,5)} = \frac{e^{i\delta}F_{25}}{2\sqrt{\xi+1}}$$

$$[\rho_s(t)]_{W_\xi}^{T_B\ (3,3)} = \frac{\xi}{2\xi+2}, [\rho_s(t)]_{W_\xi}^{T_B\ (4,1)} = \frac{F_{23}\xi e^{i(\delta-\phi)}}{2\sqrt{\xi(\xi+1)}}$$

$$[\rho_s(t)]_{W_\xi}^{T_B\ (5,2)} = \frac{e^{-i\delta}F_{25}{}^*}{2\sqrt{\xi+1}}, [\rho_s(t)]_{W_\xi}^{T_B\ (5,5)} = \frac{1}{2\xi+2}$$

$$[\rho_s(t)]_{W_\xi}^{T_B\ (7,1)} = \frac{\sqrt{\xi}e^{-i\phi}F_{35}{}^*}{2\xi+2} \quad (45)$$

This matrix is also Hermitian, and its trace norm is equal to the sum of the absolute values of its eigenvalues. According to the negativity relation, we can write:

$$N_{B-CA} = -\frac{1}{8(\xi+1)^{5/2}\sqrt{\xi(\xi+1)}}$$

$$\left(-\xi\left|\sqrt{\xi}(\xi+1)(\xi+2) - e^{-i(\delta+\phi)}\sqrt{e^{2i(\delta+\phi)}\xi(\xi+1)^2(\xi^2+4(\xi+1)|F_{25}^\mathcal{N}|^2)}\right|\right.$$

$$-\left|\sqrt{\xi}(\xi+1)(\xi+2) - e^{-i(\delta+\phi)}\sqrt{e^{2i(\delta+\phi)}\xi(\xi+1)^2(\xi^2+4(\xi+1)|F_{25}^\mathcal{N}|^2)}\right|$$

$$-\xi\left|\sqrt{\xi}(\xi+1)(\xi+2) + e^{-i(\delta+\phi)}\sqrt{e^{2i(\delta+\phi)}\xi(\xi+1)^2(\xi^2+4(\xi+1)|F_{25}^\mathcal{N}|^2)}\right|$$

$$-\left|\sqrt{\xi}(\xi+1)(\xi+2) + e^{-i(\delta+\phi)}\sqrt{e^{2i(\delta+\phi)}\xi(\xi+1)^2(\xi^2+4(\xi+1)|F_{25}^\mathcal{N}|^2)}\right|$$

$$-4\sqrt{\xi}\sqrt{\xi(\xi+1)}(\xi+1)^{3/2}\left|\sqrt{(\xi+1)|F_{23}^\mathcal{N}|^2+|F_{35}^\mathcal{N}|^2}\right| + 2\xi\sqrt{\xi(\xi+1)}(\xi+1)^{3/2}$$

$$\left. +4\sqrt{\xi(\xi+1)}(\xi+1)^{3/2}\right) \quad (46)$$

Similar to subsystems $A$ and $B$ for the subsystem $C$, the non-zero elements of the partial transpose density matrix can be expressed as follows:

$$[\rho_s(t)]_{W_\xi}^{T_C\ (1,4)} = \frac{F_{23}\xi e^{i(\delta-\phi)}}{2\sqrt{\xi(\xi+1)}}, [\rho_s(t)]_{W_\xi}^{T_C\ (1,6)} = \frac{e^{i\delta}F_{25}}{2\sqrt{\xi+1}}$$

$$[\rho_s(t)]_{W_\xi}^{T_C\ (2,2)} = 1/2, [\rho_s(t)]_{W_\xi}^{T_C\ (3,3)} = \frac{\xi}{2\xi + 2}$$

$$[\rho_s(t)]_{W_\xi}^{T_C\ (3,5)} = \frac{F_{35}\sqrt{\xi}e^{i\phi}}{2\xi + 2}, [\rho_s(t)]_{W_\xi}^{T_C\ (4,1)} = \frac{\xi F_{23}{}^* e^{-i(\delta-\phi)}}{2\sqrt{\xi(\xi+1)}}$$

$$[\rho_s(t)]_{W_\xi}^{T_C\ (5,3)} = \frac{\sqrt{\xi}e^{-i\phi}F_{35}{}^*}{2\xi + 2}, [\rho_s(t)]_{W_\xi}^{T_C\ (5,5)} = \frac{1}{2\xi + 2}$$

$$[\rho_s(t)]_{W_\xi}^{T_C\ (6,1)} = \frac{e^{-i\delta}F_{25}{}^*}{2\sqrt{\xi+1}} \quad (47)$$

This matrix is also Hermitian. The trace norm is the sum of the absolute values of its eigenvalues. Following the negativity relation, we can deduce the following:

$$N_{C-AB} = -\frac{1}{8(\xi+1)^{3/2}\sqrt{\xi(\xi+1)}}$$

$$(-\left|\sqrt{\xi}(\xi+1)^2 - e^{-i(\delta+\phi)}\sqrt{e^{2i(\delta+\phi)}\xi(\xi+1)^2\left(4\xi|F_{35}^{\mathcal{N}}|^2 + (\xi-2)\xi + 1\right)}\right|$$

$$-\left|\sqrt{\xi}(\xi+1)^2 + e^{-i(\delta+\phi)}\sqrt{e^{2i(\delta+\phi)}\xi(\xi+1)^2\left(4\xi|F_{35}^{\mathcal{N}}|^2 + (\xi-2)\xi + 1\right)}\right|$$

$$-4\xi\sqrt{\xi(\xi+1)}\left|\sqrt{\xi|F_{23}^{\mathcal{N}}|^2 + |F_{25}^{\mathcal{N}}|^2}\right| - 4\sqrt{\xi(\xi+1)}\left|\sqrt{\xi|F_{23}^{\mathcal{N}}|^2 + |F_{25}^{\mathcal{N}}|^2}\right|$$

$$+2\sqrt{\xi(\xi+1)}(\xi+1)^{3/2}) \quad (48)$$

## 6 Results and discussion

Quantum entanglement is crucial for tasks in quantum information and computation [12, 13, 19, 17, 10]. Spin systems are known for their suitability for quantum information processing, and various studies have explored their entanglement and discord properties [19, 17, 10]. However,

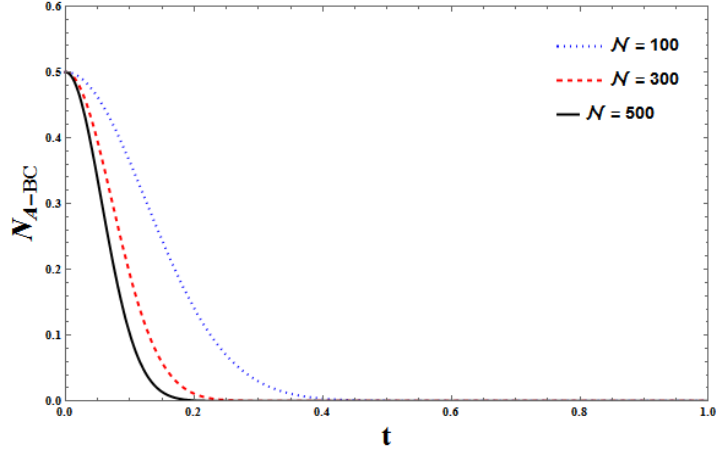

Figure 1: The negativity diagram according to the time for the $GHZ$ state and the three subsystems $A$, $B$ and $C$ corresponding to it for certain values of $g^A = 0.1$, $g^B = 0.2$, $g^C = 0.5$, $h_i = 1$ and $\gamma_i = \eta_i = 1/\sqrt{2}$, for three different values of $\mathcal{N}$, $\mathcal{N} = 100$, $\mathcal{N} = 300$, $\mathcal{N} = 500$.

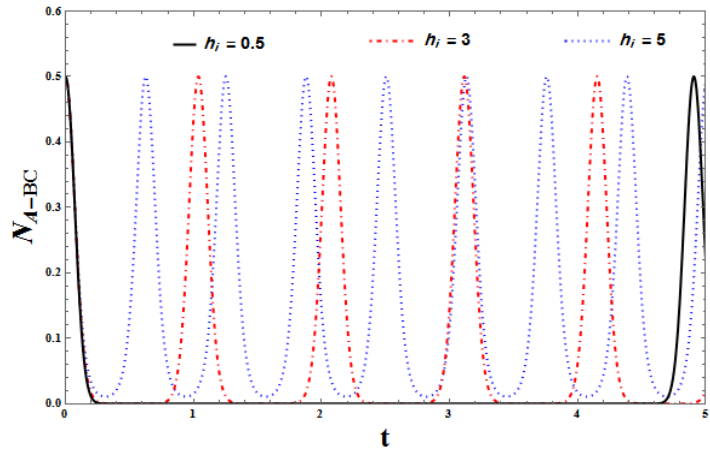

Figure 2: Negativity diagram according to the time for $GHZ$ state and three subsystems $A$, $B$ and $C$ corresponding to it, for specific values of $g^A = 0.1$, $g^B = 0.2$, $g^C = 0.5$, $\mathcal{N} = 300$, and $\gamma_i = \eta_i = 1/\sqrt{2}$, for three different values of $h_i$, $h_i = 0.5$, $h_i = 3$, $h_i = 5$.

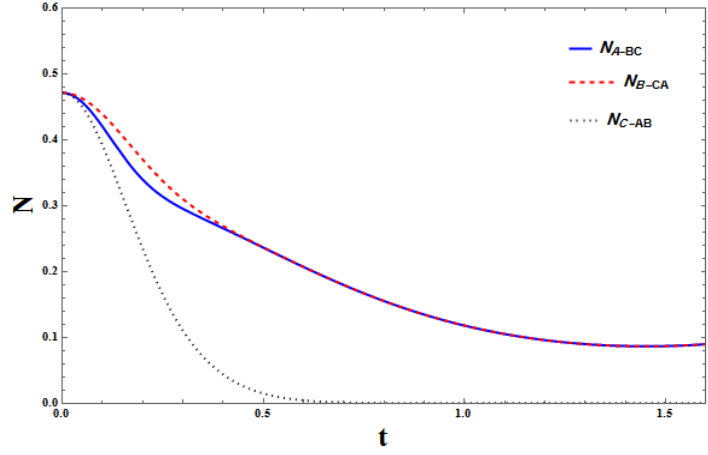

Figure 3: Negativity diagram according to the time for state $W$ and three subsystems $A$, $B$ and $C$ corresponding to it for certain values $g^A = 0.1$ and $g^B = 0.2$ and $g^C = 0.5$ and $\mathcal{N} = 300$ and $\gamma_i = \eta_i = 1/\sqrt{2}$ and $h_i = 1$.

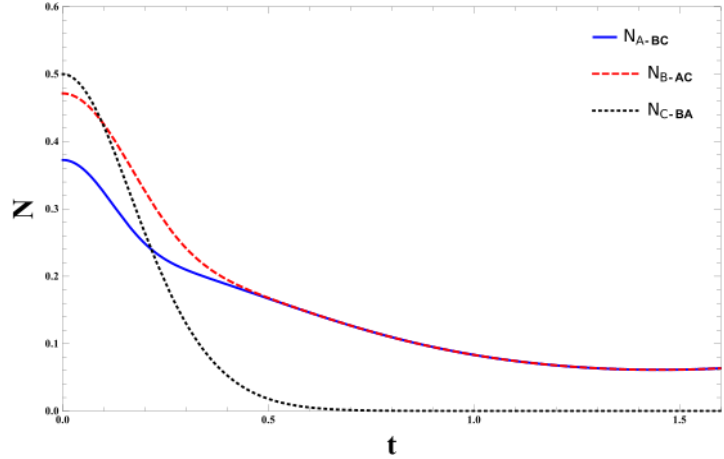

Figure 4: The negativity diagram according to the time for $W_\xi$ state and three subsystems $A$, $B$ and $C$ corresponding to it, for specific values of $g^A = 0.1$, $g^B = 0.2$, $g^C = 0.5$, $\mathcal{N} = 300$, and $\gamma_i = \eta_i = 1/\sqrt{2}$ and $h_i = 1$, $\delta = \phi = 0$ and $\xi = 2$.

These quantum correlations are vulnerable to decoherence caused by interactions with the environment [18, 16, 14]. Although the entanglement dynamics in spin systems under decoherence from different qubit environments have been studied [9, 6, 11], the effects of decoherence from qutrit environments have not been explored.

We begin by presenting the results for the $GHZ$ state as the initial state of the three-qubit system. Building on the topics discussed in the preceding sections, the negativity relationship remains consistent across the three subsystems $A$, $B$ and $C$ and can be expressed as follows:

$$N_{A-BC} = N_{B-CA} = N_{C-AB} \quad (49)$$

Fig. 1 illustrates the evolution of negativity for the $GHZ$ state over time with three different $\mathcal{N}$ values. It can be observed that the negativity diminishes after a certain duration; however, as the number of particles in the spin chain increases, the decay occurs more rapidly.

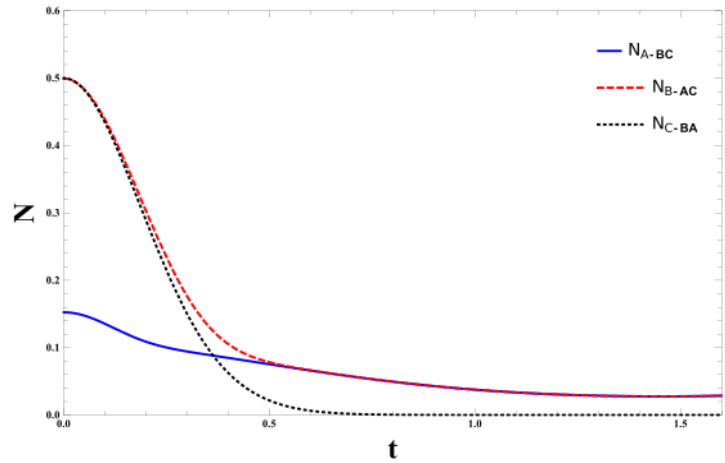

Figure 5: The negativity diagram according to the time for $W_\xi$ state and three subsystems $A$, $B$ and $C$ corresponding to it, for specific values $g^A = 0.1$, $g^B = 0.2$, $g^C = 0.5$, $\mathcal{N} = 300$ and $\gamma_i = \eta_i = 1/\sqrt{2}$ and $h_i = 1$, $\delta = \phi = 0$ and $\xi = 20$.

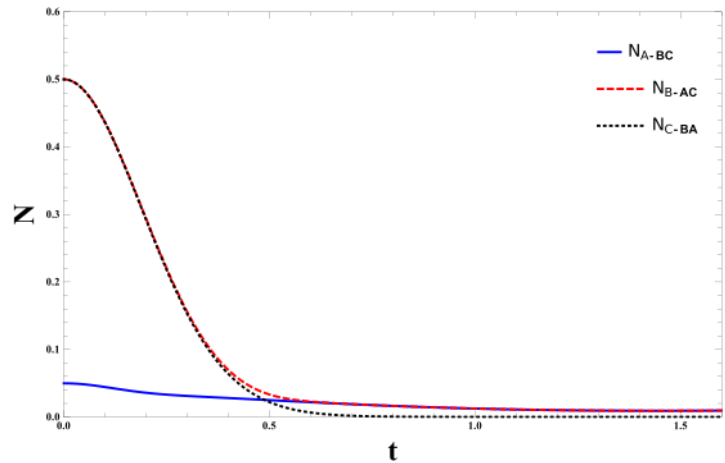

Figure 6: Negativity diagram according to the time for $W_\xi$ state and three subsystems $A$, $B$ and $C$ corresponding to it, for the known values of $g^A = 0.1$, $g^B = 0.2$, $g^C = 0.5$, $\mathcal{N} = 300$, and $\gamma_i = \eta_i = 1/\sqrt{2}$ and $h_i = 1$, $\delta = \phi = \pi/4$ and $\xi = 200$.

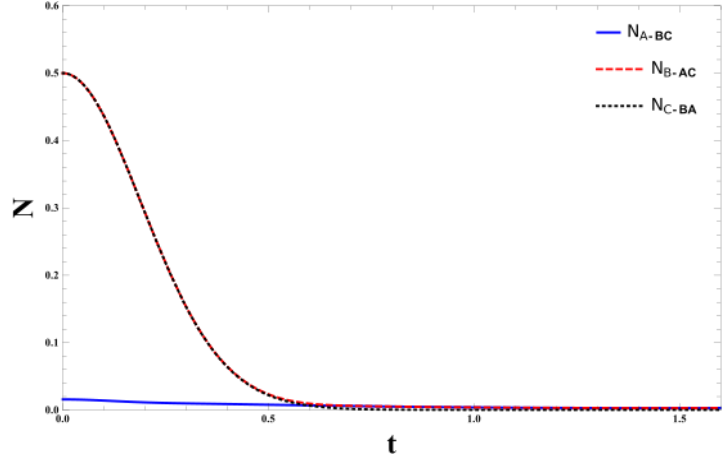

Figure 7: Negativity diagram according to the time for $W_\xi$ state and three subsystems $A$, $B$ and $C$ corresponding to it, for the known values of $g^A = 0.1$, $g^B = 0.2$, $g^C = 0.5$, $\mathcal{N} = 300$, and $\gamma_i = \eta_i = 1/\sqrt{2}$ and $h_i = 1$, $\delta = \phi = \pi/4$ and $\xi = 2000$.

Fig. 2, we depict the negativity evolution for the $GHZ$ state over time with three different $h_i$ values. This figure shows that the negativity exhibits an oscillatory behavior resembling sinusoidal evolution. A higher $h_i$ value results in a shorter period. Moreover, Fig. 2 indicates that entanglement disappears at low $h_i$ values, and the end time of entanglement extending as $h_i$ decreases. The negativity relationship remains consistent for the three subsystems $A$, $B$ and $C$ in the $GHZ$ state. Fig. 3 presents the evolution of negativity over time for the $W$ state. This figure shows that the negativity decreases over time, specifically for the $W$ state, with the loss of entanglement observed only for $N_{C-AB}$.

Fig. 4 and Fig. 5 illustrates the time evolution of the negativity for the $W_\xi$ state with $\delta = \phi = 0$, $\xi = 2$, and $\xi = 20$. Meanwhile, Fig. 6 and Fig. 7 displays the time evolution of the negativity for the $W_\xi$ state with $\delta = \phi = \pi/4$, $\xi = 200$ and $\xi = 2000$. Throughout Fig. 4 to Fig. 7, the negativity decreases over time for the $W_\xi$ state. As $\xi$ increases, $N_{B-AC}$ and $N_{C-AB}$ converge. Consequently, at very high $\xi$ values, entanglement ceases for these pairs. At the same time, $N_{A-BC}$ also shows a significant reduction in entanglement for substantial $\xi$ values.

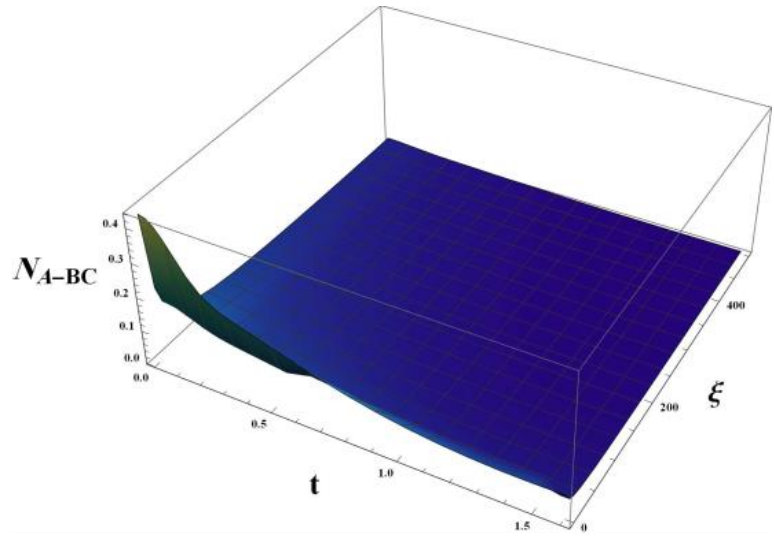

Figure 8: The graph of $N_{A-BC}$ for $W_\xi$ state, in $\xi$ and $t$ for certain values of $\delta = \phi = \pi/4$.

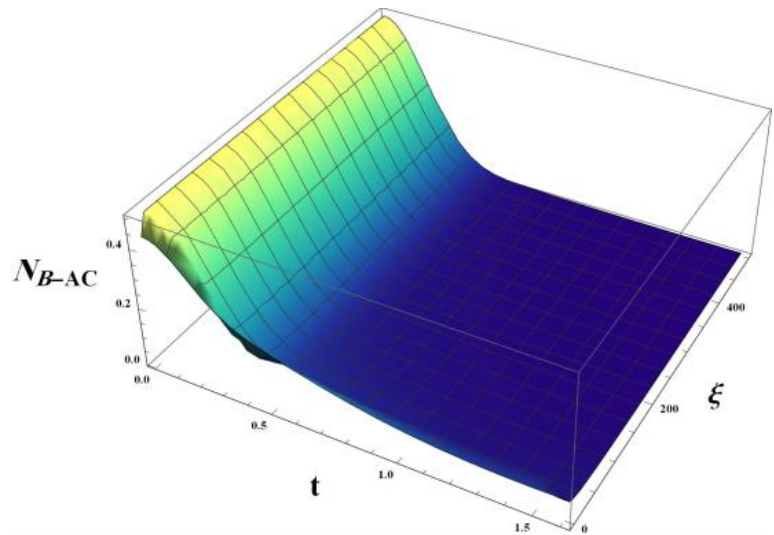

Figure 9: The graph of $N_{B-AC}$ for $W_\xi$ state, in $\xi$ and $t$ for certain values of $\delta = \phi = \pi/4$.

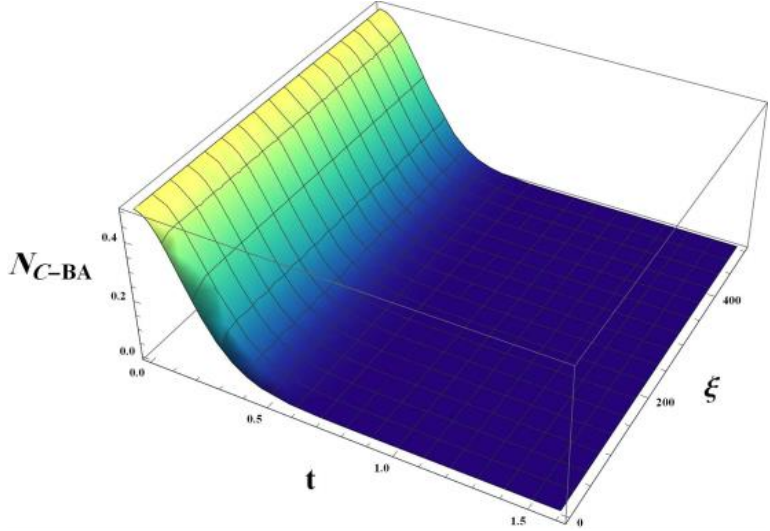

Figure 10: The graph of $N_{C-BA}$ for $W_\xi$ state, in $\xi$ and $t$ for certain values of $\delta = \phi = \pi/4$.

Figs. 8, Fig. 9, and Fig. 10 depict the $N_{A-BC}$, $N_{B-AC}$ and $N_{C-AB}$ diagrams for the state $W_\xi$ in $\xi$ and $t$. Overall, the analysis of Fig. 8 to Fig. 10 indicates that as $\xi$ increases over time, $N_{B-AC}$ and $N_{C-AB}$ remain relatively stable, whereas $N_{A-BC}$ decreases notably with $\xi$ growth.

The in-depth research outcomes detailed in this article are closely aligned with the data obtained from the investigations by Guo et al. [5], highlighting a strong correlation between the two sets of findings.

Quantum entanglement dynamics in a three-qubit system interacting with a spin chain hold significance because they provide a deeper understanding of how various quantum states correlate with one another and how they can influence each other's behavior in intricate ways. By examining this interaction, researchers can gain valuable insights that contribute to the field of quantum information processing, which is essential for the outcome of advanced quantum computing techniques. In addition, studying these dynamics can improve the overall efficiency of quantum computing systems, thus making them more powerful and effective. In addition, this exploration enhances our fundamental understanding of quantum phenomena, which are often counterintuitive and complex. Ultimately, such advancements can pave the way for innovative developments within the realm of quantum technologies, opening new avenues for both practical applications and theoretical research. The quantum entanglement dynamics in a three-qubit system interacting with a spin chain contribute to the broader field by providing insights into how entangled quantum states behave and interact. This understanding is essential for advancing quantum information theory, improving quantum computing protocols, and developing new quantum technologies. By studying these dynamics, researchers can explore the fundamental principles of quantum mechanics and potentially discover new applications in areas like quantum communication, cryptography, and materials science.

## 7  Conclusion

This study investigates the intricate evolution of entanglement in three-qubit states as they interact with a spin-chain environment. We use negativity as a key metric for measuring entanglement, which plays a vital role in the analysis of dynamic behavior. Our investigation focuses on three notable states: the $GHZ$ state, the $W$ state, and the $W_\xi$ state, all of which take into account the proposed Hamiltonian that guides our analysis.

Our findings indicate that the negativity associated with the $GHZ$ state experiences a gradual decrease over time, occurring at an accelerated rate as the number of particles within the system increases. Furthermore, the time evolution of negativity exhibits an oscillatory pattern influenced by varying values of $h_i$, suggesting complex dynamics.

Similarly, the $W$ state also undergoes a noticeable reduction in negativity as time progresses, reflecting changes in its entanglement characteristics.

In the case of the $W_\xi$ state, we observe that negativity continues to decrease with time, and particularly interesting is the phenomenon of entanglement death that emerges in the $A$ subsystem when the parameter $\xi$ is increased. This observation highlights the delicate balance between entanglement and the system. Our analysis reveals that the negativity measures $N_{B-CA}$ and $N_{C-AB}$ remain relatively stable throughout the process, whereas $N_{A-BC}$ experiences a significant decline as $\xi$ continues to increase.

Overall, the results obtained are consistent with findings from similar studies in the field, demonstrating a strong alignment with previous research outcomes. This consistency reinforces the validity of our conclusions and contributes to a deeper understanding of quantum entanglement dynamics in complex systems.

The field of quantum entanglement dynamics in various systems is a vibrant research area that is constantly evolving, and significant progress is being made on several fronts. The discovery of quantum entanglement (QE) dynamics is particularly important in theoretical and practical applications, especially in quantum information science and quantum computing. Learning these dynamics typically involves understanding unitary transformations and observables in quantum systems. Researchers have focused on predicting the outcomes of quantum operations using machine learning techniques. Specifically, the challenge lies in approximating functions that express the effect of quantum operations on state vectors, which involves using training datasets of quantum states and their respective measurement outputs. A recent study discussed relevant hypotheses that minimize prediction errors based on entangled data, thereby highlighting how entangled datasets can provide quantum advantages in learning tasks. Research has increasingly recognized the distinction between closed and open quantum systems. Open quantum systems interact with their environments, resulting in decoherence phenomena. The dynamics of quantum entanglement can differ significantly from those of closed systems, influencing the coherence of entangled states over time. Specifically, researchers are interested in studying non-Markovian dynamics, where a system's past state influences its future states, in contrast to memory-less Markovian interactions [15]. Quantum entanglement is a critical resource for various applications such as

Quantum Computing: Qubits, which are the basic units of quantum information, exploit superposition and entanglement to perform computations that classical bits cannot efficiently handle. The potential of quantum computers to outperform classical computers lies in these principles [7].

Quantum Teleportation: This process involves transmitting quantum states between distant particles via entanglement. The current research focuses on enhancing the effectiveness of teleported states driven by the specifics of entanglement dynamics [4].